\documentclass[12pt]{article}

\usepackage[T1]{fontenc}
\usepackage[utf8]{inputenc}
\usepackage{amsmath}                 % BEFORE newtxmath
\usepackage{newtxtext,newtxmath}

\usepackage[letterpaper,margin=1in]{geometry}
\usepackage{graphicx}
\usepackage{upgreek}
\usepackage{xcolor}
\usepackage{needspace}
\usepackage{url}

\usepackage{scicite}
\usepackage{cite}

\usepackage{xr-hyper}
\usepackage[colorlinks=true, allcolors=blue]{hyperref}
\usepackage{breakurl}

\usepackage{soul}

\date{}

\renewenvironment{abstract}{\quotation}{\endquotation}
\makeatletter
\renewcommand{\fnum@figure}{\textbf{Figure \thefigure}}
\renewcommand{\fnum@table}{\textbf{Table \thetable}}
\makeatother

\def\scititle{
	X-ray driven displacive excitation of coherent phonons
}
\title{\bfseries \boldmath \scititle}

\author{
    Oleg~Dogadov$^{1,2}$,
    Remi~Claude$^{3}$,
    Christoph~Emeis$^{4}$,
    Giovanni~Batignani$^{5,6}$,\and
    Giuseppe~Fumero$^{5\dagger}$,
    Roberto~Costantini$^{7,8}$,
    Agata~Azzolin$^{9,10,11}$,\and
    Sabine~Rockenstein$^{9,10,12}$,
    Paolo~Cattaneo$^{3}$,
    Rebeca Gomez~Castillo$^{13}$,\and
    Matteo~Manzi$^{13}$,
    Angelo~Giglia$^{7}$,
    Dario De~Angelis$^{14}$,\and
    Ettore~Paltanin$^{14}$,
    Zeinab~Ebrahimpour$^{14}$,
    Marija~Krstulovic$^{14}$,
    Gabor~Kurdi$^{14}$,\and
    Miltcho~Danailov$^{14}$,
    Luca~Giannessi$^{14,15}$,
    Riccardo~Mincigrucci$^{14}$,\and
    Emiliano~Principi$^{14}$,
    Alberto~Crepaldi$^{1}$,
    Giulio~Cerullo$^{1,16}$,\and
    Giovanni De~Ninno$^{14,17}$,
    Claudio~Masciovecchio$^{14}$,
    Laura~Foglia$^{14}$,\and
    Tullio~Scopigno$^{5,6,18}$,
    Fabrizio~Carbone$^{3,13}$,
    Francesca~Calegari$^{9,10,11}$,\and
    Fabio~Caruso$^{4}$,
    Michele~Puppin$^{13}$,
    Oliviero~Cannelli$^{9,11,13*}$\and
	\small$^{1}$Dipartimento di Fisica, Politecnico di Milano, Piazza Leonardo da Vinci, 32, 20133 Milano, Italy.\and
	\small$^{2}$Department of Physical Chemistry, Fritz Haber Institute of the Max Planck Society, \and \small 14195 Berlin, Germany.\and
	\small$^{3}$Laboratory for Ultrafast Microscopy and Electron Scattering (LUMES), Institute of Physics, \and \small École Polytechnique Fédérale de Lausanne (EPFL), Lausanne 1015 CH, Switzerland.\and
    \small$^{4}$Institute of Theoretical Physics and Astrophysics, Kiel University, 24118 Kiel, Germany.\and
    \small$^{5}$Dipartimento di Fisica, Sapienza Università di Roma, Roma, Italy.\and
    \small$^{6}$Istituto Italiano di Tecnologia (IIT), Center for Life Nano Science @Sapienza, Roma, Italy.\and
    \small$^{7}$CNR—Istituto Officina dei Materiali (IOM), S.S. 14 km 163.5, 34149 Trieste, Italy.\and
    \small$^{8}$Dipartimento di Fisica, Università di Trieste, Via Valerio 2, I-34127 Trieste, Italy.\and
    \small$^{9}$Center for Free-Electron Laser Science, DESY, Notkestraße 85, 22607 Hamburg, Germany.\and
    \small$^{10}$Physics Department, University of Hamburg, Luruper Chausee 149, 22761 Hamburg, Germany.\and
    \small$^{11}$The Hamburg Centre for Ultrafast Imaging, University of Hamburg, Luruper Chausee 149, \and \small 22761 Hamburg, Germany.\and
    \small$^{12}$Max-Planck-Institut für Struktur und Dynamik der Materie, Hamburg, Germany.\and
    \small$^{13}$Lausanne Centre for Ultrafast Science (LACUS), École Polytechnique Fédérale de Lausanne (EPFL), \and \small  CH-1015 Lausanne, Switzerland.\and
    \small$^{14}$Elettra-Sincrotrone Trieste S.C.p.A., Strada Statale 14, km 163.5, AREA Science Park, \and \small I-34149, Basovizza, Trieste, Italy.\and
    \small$^{15}$Istituto Nazionale di Fisica Nucleare, Laboratori Nazionali di Frascati, 00044 Frascati, Italy.\and
    \small$^{16}$CNR-IFN, Piazza Leonardo da Vinci 32, 20133 Milan, Italy.\and
    \small$^{17}$Laboratory of Quantum Optics, University of Nova Gorica, Si-5270 Ajdovščina, Slovenija.\and
    \small$^{18}$Istituto Italiano di Tecnologia (IIT), Graphene Labs, Genova, Italy.\and
    \small$^\dagger$Present address: CNR NANOTEC, Institute of Nanotechnology, Via Monteroni, Lecce 73100, Italy.\and
	\small$^\ast$Corresponding author. Email: oliviero.cannelli@cfel.de
}

\begin{document} 

% Insert the title and author list
\maketitle

\begin{abstract} \bfseries \boldmath
% Science abstract:
Modulating electron–phonon coupling offers a route to control structural displacements and tune material functionality. Valence-to-conduction band transitions, however, provide limited leverage over the driving force.
Here, we demonstrate coherent lattice dynamics in trigonal tellurium using free-electron laser pulses tuned to the Te $\mathrm{N_{4,5}}$-edge. Over a broad fluence range, the oscillation amplitude obeys the displacive excitation of coherent phonons framework, extended to core resonance with twice the driving efficiency of a visible pump.
\textit{Ab initio} calculations decompose the force into competing multiband contributions, inaccessible to optical excitation, whose balance shifts as carriers relax. Tunable extreme-ultraviolet and X-ray pulses thus open a regime in which the displacive response is set by band-dependent coupling to the lattice, not by the number and temperature of the photocarriers alone.

% Science Advances abstract:
% Modulating electron–phonon coupling offers a route to selectively drive lattice displacements and tune material functionality. Most ultrafast studies, however, rely on optical excitation of valence-to-conduction band transitions, providing limited leverage over the microscopic origin of the driving force. Here, we experimentally demonstrate coherent lattice dynamics in trigonal tellurium initiated by ultrashort extreme-ultraviolet pulses from a free-electron laser tuned to the Te N$_{4,5}$ core transition. A shot-to-shot chirped detection scheme enables measurements of the coherent $\mathrm{A_1}$ phonon over a broad excitation density range. Its amplitude obeys the displacive excitation of coherent phonons framework, extending it to core-resonant excitation with twice the driving efficiency of a visible pump. \textit{Ab initio} calculations decompose the force into  multiband contributions from both electrons and shallow-core holes, the latter inaccessible to optical pumping, whose balance shifts as carriers relax. Tunable extreme-ultraviolet and X-ray pulses thus open a regime in which the displacive response is set by band-dependent coupling to the lattice, not by the number and temperature of the photocarriers alone.

\end{abstract}

%%%%%%%%%%%%%%%%%%%%%%%%%%%%%%%%%%%%%%%%%%%%%%%%%%%%%%%%
%%%%%%%%%%%%%%%%%%%%%%%%%%%%%%%%%%%%%%%%%%%%%%%%%%%%%%%%
%%%%%%%%%%%%%%%%%%%%%%%%%%%%%%%%%%%%%%%%%%%%%%%%%%%%%%%%
%%%%%%%%%%%%%%%%%%%%%%%%%%%%%%%%%%%%%%%%%%%%%%%%%%%%%%%%

\noindent

\subsection*{Introduction}

Controlling functionality in materials through selective light excitations is a long-standing goal of ultrafast science. Pioneering works leveraged near-infrared (NIR) and visible femtosecond laser pulses to activate Raman-active modes through resonant and non-resonant impulsive excitation of crystalline systems \cite{cheng_mechanism_1991,dhar1994time,merlin1997generating,stevens2002coherent}. This process, mediated by electron-phonon coupling (EPC), established the foundation of coherent phonon spectroscopy in condensed phase \cite{cerullo2000femtosecond,kahan2007following,ishioka2009coherent,liebel2015principles,batignani2018probing} and enabled displacive lattice dynamics to be used as a tool to probe and manipulate electronic structure and structural phase. Representative achievements include non-thermal insulator-to-metal phase transition in vanadium oxides and rare-earth nickelates \cite{wall2012ultrafast, johnson2022ultrafast, dogadov2025ultrafast}, coherent control of chemical potential in $\mathrm{BaFe_2As_2}$ \cite{yang2014ultrafast}, light-induced Weyl semiconductor-to-metal phase transition in tellurium \cite{ning2022light}, photoinduced multistage phase transition in $\mathrm{Ta_2NiSe_5}$ \cite{liu2021photoinduced}, and vibrational coherent control over phase-transitions in transition-metal dichalcogenides \cite{maklar2023coherent,sayers2024mapping}.

% Despite this success, existing strategies offer limited tunability of the EPC-mediated displacive driving force because NIR and visible excitation are restricted to valence-to-conduction band electronic transitions. Shorter-wavelength excitation at free-electron lasers (FELs) accesses core-level transitions with intrinsic element and orbital specificity \cite{gahl2008femtosecond,ferrer2015nonlinear,chergui2023progress}. However, coherent phonon activation at FELs has so far been reported only in nonlinear experiments, such as transient grating (TG) under off-resonant excitation \cite{bencivenga2015four,maznev2018generation,rouxel2021hard,ferrari2025all}. Whether core-level resonant excitation can impulsively generate coherent phonons and how it connects to the established displacive excitation of coherent phonons (DECP) framework \cite{zeiger_theory_1992,merlin1997generating} has remained largely unexplored.

Despite this success, existing strategies offer limited tunability of the EPC-mediated displacive driving force because NIR and visible excitations are restricted to valence-to-conduction band electronic transitions. Shorter-wavelength excitation with free-electron lasers (FELs) accesses core-level transitions with intrinsic element and orbital specificity \cite{gahl2008femtosecond,ferrer2015nonlinear,chergui2023progress}. 
% However, coherent phonon activation at FELs has so far been reported in nonlinear experiments such as transient grating (TG) under off-resonant excitation \cite{bencivenga2015four,maznev2018generation,rouxel2021hard,ferrari2025all}, which is mediated by spatial photocarrier and thermal transport contributions. 
Coherent phonon activation at FELs has so far been achieved through off-resonant nonlinear four-wave-mixing schemes, such as transient grating (TG) spectroscopy, in which two crossed beams imprint a spatially periodic excitation at finite wavevector \cite{bencivenga2015four,maznev2018generation,rouxel2021hard,ferrari2025all}.
Whether core-level resonant excitation can impulsively generate phonons and how it connects to the established displacive excitation of coherent phonons (DECP) framework \cite{zeiger_theory_1992,merlin1997generating} has remained largely unexplored.
Equally open is whether the driving force depends only on the number of carriers promoted and their temperature \cite{drescher2025onset}, as in the original DECP formulation, or also on the electronic states they occupy.

Trigonal tellurium provides a well-established platform to address this question. It is a Peierls-distorted small band gap semiconductor \cite{tutihasi_optical_1969} with a unit cell containing three covalently-bonded non-equivalent sites, which forms helical chains that are aligned along the crystallographic \textit{c} axis and interact through van der Waals forces. The transient response of tellurium upon excitation with a femtosecond laser pulse is dominated by the fully symmetric $\mathrm{A_1}$ mode (Figure~\ref{fig:main}A), making it a benchmark system for studying DECP dynamics \cite{cheng_mechanism_1991,hunsche_impulsive_1995,kudryashov2007intraband,kamaraju_large-amplitude_2010,dekorsy1995emission,huber2015coherent,gatti2024light}. In the present work, extreme-ultraviolet (XUV) pulses from the FERMI FEL \cite{allaria_fermi_2015} are tuned just above the Te N$_{4,5}$ absorption edge to promote an ultrafast core-to-conduction band transition. The reflectivity oscillations due to the excited coherent $\mathrm{A_1}$ phonon are detected as a function  of time delay using a NIR probe (Figure~\ref{fig:main}B).

A key aspect of this approach is the implementation of chirped-probe detection with shot-to-shot acquisition at a FEL facility. By encoding temporal information in the probe spectrum, this method enables excitation density dependent measurements that would otherwise require prohibitively long acquisition times with conventional delay scans or would increase the risk of damage. This allows a direct test of the excitation mechanism, demonstrating that the phonon response follows the linear scaling expected for DECP, but with a twofold higher driving efficiency than optical pumps, establishing X-ray driven displacive excitation of coherent phonons (XDECP).

% % Building on this experimentally established framework, 
% We combine our measurements with out-of-equilibrium \textit{ab initio} simulations that extend the displacive excitation description to include multiband carrier dynamics following core-resonant excitation.
% The calculations reveal how distinct electronic bands contribute with competing terms to the time-dependent electron–phonon force during ultrafast carrier relaxation. These results show that, although the observed lattice vibration is analogous to that launched optically, its microscopic activation pathway depends sensitively on how the initial electronic configuration is prepared. 
% %Core-resonant excitation therefore offers a route to tailor non-equilibrium carrier distributions and new strategies for engineering the displacive driving force in solids.
% Core-resonant excitation therefore offers new strategies for engineering the displacive driving force in solids by selectively tailoring non-equilibrium carrier distributions and the associated band-specific EPC contributions to the lattice dynamics.

% Building on this experimentally established framework, 
Combined with out-of-equilibrium \textit{ab initio} simulations, our approach extends the displacive excitation description to include multiband carrier dynamics under core-level pumping.
% Specifically, the calculation reveals how distinct electronic bands contribute with competing terms to the time-dependent electron–phonon force during ultrafast carrier relaxation. These results show that, although the observed lattice vibration is analogous to that launched optically, its microscopic activation pathway depends sensitively on how the initial electronic configuration is prepared. 
Specifically, the calculation reveals that shallow-core holes and conduction band electrons contribute to the electron–phonon force with opposite signs, and that their contributions reach a maximum at different stages of the carrier relaxation. The lattice vibration that results is analogous to the one launched optically, but the force that drives it is not.
Core-resonant excitation therefore offers new strategies for engineering the displacive driving force in solids by selectively tailoring non-equilibrium carrier distributions and the associated band-specific EPC contributions to the lattice dynamics.

%%%%%%%%%%%%%%%%%%%%%%%%%%%%%%%%%%%%%%%%%%%%%%%%%%%%%%%%
%%%%%%%%%%%%%%%%%%%%%%%%%%%%%%%%%%%%%%%%%%%%%%%%%%%%%%%%
%%%%%%%%%%%%%%%%%%%%%%%%%%%%%%%%%%%%%%%%%%%%%%%%%%%%%%%%
%%%%%%%%%%%%%%%%%%%%%%%%%%%%%%%%%%%%%%%%%%%%%%%%%%%%%%%%

\noindent

\subsection*{Results}

\begin{sloppypar}
\paragraph{Experimental demonstration of core-resonant displacive dynamics:}

Figure~\ref{fig:main}C shows a schematic representation of the electronic transitions induced by the 46.2~eV pump at the high-energy side of the Te N$_{4,5}$-edge (see X-ray absorption spectrum in Figure~\ref{fig:XAS} of the Supplementary Material, SM). The temporal evolution of the room temperature transient reflectivity signal probed in the 1.56--1.59~eV energy range upon core-resonant excitation is reported in Figure~\ref{fig:main}D, evidencing uniform coherent lattice motion that is independent of the probe photon energy. 
The map was corrected for photon energy-dependent time zero due to the employed chirped detection (see section \ref{SM:XUV_analysis} of the SM for details).
The lower panel shows the spectrally averaged time trace up to 4.2~ps. The response consists of an incoherent electronic component superimposed on a coherent vibrational contribution, which we separate using the model
\end{sloppypar} % this is needed to avoid that the name of the paragraph exceeds the line lenght in the pdf
%
%\hl{Core-resonant excitation at the Te N$_{4,5}$-edge generates a pronounced oscillatory modulation of the transient reflectivity, evidencing coherent lattice motion}. Figure~\ref{fig:main}c shows a diagrammatic representation of the electronic transitions (see X-ray absorption spectrum in Figure~\ref{fig:XAS} of the \hyperref[supplementary material]{Supplementary Material}, SM), while the temporal evolution of the chirp-corrected transient reflectivity signal in the 1.56--1.59~eV energy range is reported in Figure~\ref{fig:main}d. The lower panel shows the spectrally averaged time trace up to 4.2~ps. The response consists of an incoherent electronic component superimposed on a coherent vibrational contribution, which we separate using the model
%
% {\small
% \begin{equation}
% \begin{aligned}
% \Delta R(t)/R = &
% A \cdot \exp(-t{/\tau_\mathrm{el}}) + \\
% & A_\mathrm{osc} \cdot \cos \bigl( 2\pi\nu_{ph} t + \phi \bigr) \cdot \exp(-{\lambda}\cdot t) +
% c,
% \end{aligned}
% \label{eq:transients}
% \end{equation}
% }
%
\begin{equation}
\Delta R(t)/R = 
A \cdot \exp(-t{/\tau_\mathrm{el}}) +
A_\mathrm{osc} \cdot \cos \bigl( 2\pi\nu_\mathrm{ph} t + \phi \bigr) 
\cdot \exp(-t/\tau_{damp}) + c,
\label{eq:transients}
\end{equation}
where $A$ and $A_\mathrm{osc}$ are the amplitudes of the non-oscillatory and oscillatory components, respectively, and $c$ is a constant offset. From the spectrally-averaged transient reflectivity trace we extract the decay time of the non-oscillatory component $\tau_\mathrm{el} = 2.0 \pm 0.2$~ps, a phonon frequency $\nu_\mathrm{ph} = 3.535 \pm 0.003$~THz consistent with the $\mathrm{A_1}$ coherent mode of trigonal tellurium \cite{cheng_mechanism_1991,hunsche_impulsive_1995,kudryashov2007intraband,kamaraju_large-amplitude_2010,dekorsy1995emission,huber2015coherent,gatti2024light}, and a damping time $\tau_{damp} = 1.8 \pm 0.1$~ps.

To determine whether this excitation follows a displacive mechanism or arises from nonlinear FEL-driven processes such as TG \cite{chergui2023progress}, we measure the response as a function of excitation fluence. Figure~\ref{fig:fluence}A reports the oscillatory component of the transient reflectivity signal for FEL fluences in the range 0.08--0.42~mJ$\cdot$cm$^{-2}$. 
The measurements are enabled by a shot-to-shot chirped-probe detection scheme similar to the chirped impulsive stimulated Raman scattering developed in the optical domain \cite{batignani_broadband_2019}. 
This method assumes that the sample response is independent of probe photon energy (see Figure~\ref{fig:main}D) and uses a strongly chirped probe pulse to encode time delay into the probe spectrum, as detailed in section \ref{SM:XUV_fluences} of the SM. 
By using a linearly chirped NIR probe, we collect coherent responses spanning 0.71~ps on a single-shot basis, bypassing the stochastic noise arising from pulse-to-pulse FEL intensity fluctuations, which is a critical bottleneck in conventional FEL-pump--optical-probe scans. 
The solid lines in Figure~\ref{fig:fluence}A correspond to best fits obtained with an exponentially damped cosine function, %, $(\Delta R(t)/R)_\mathrm{osc} = A_\mathrm{osc} \cdot \cos \bigl( 2\pi\nu_\mathrm{ph} t + \phi \bigr) \cdot \exp(-{\lambda}\cdot t)$.
with the main fit parameters reported in  Figure~\ref{fig:fluence}B. The amplitude of the oscillatory component follows a linear dependence over the entire fluence range (Figure~\ref{fig:fluence}B, left axis), demonstrating the expected scaling for a displacive excitation, thereby excluding other nonlinear effects stimulated by the XUV pump \cite{tamasaku_x-ray_2014,bencivenga_four-wave_2015,rouxel_hard_2021,chergui_progress_2023,bencivenga2023extreme}. The oscillation frequency exhibits a linear softening from 3.54~THz to 3.36~THz across the investigated fluence range.
The results are compatible with NIR-pump studies, where softening \cite{hunsche_impulsive_1995,adelman2025coherently} and chirping \cite{kamaraju_large-amplitude_2010} of the $\mathrm{A_1}$ phonon frequency were attributed to anharmonic phonon--phonon coupling.
However, differently from the optical regime, the core-resonant excitation drives a significantly stronger phonon response. We quantify these changes by performing complementary time-resolved measurements using 3.10~eV pump and 1.55~eV probe pulses (see section \ref{SM:comparison_XUV_opt_fluences} of the SM for details).
Due to the intrinsic complexity in comparing absolute fluences in table-top and FEL experiments, a more compelling determination of the phonon activation strength is obtained by monitoring the amplitude of the coherent response against the average incoherent transient reflectivity signal (Figure~\ref{fig:fluence}C). In fact, both quantities scale linearly over a wide pump fluence range (see Figure \ref{fig:opt_vs_XUV_fluence}), 
providing an internal calibration that does not rely on the nominal values of the absorbed fluence and photocarrier density.
The coherent phonon driving efficiency, here defined as the slope of the linear fit between these two quantities, corresponds to \mbox{0.46 $\pm$ 0.02} and \mbox{0.23 $\pm$ 0.01} for the 46.2~eV and for the 3.1~eV pump photon energies, respectively. These results confirm that XDECP in tellurium is \mbox{1.97 $\pm$ 0.08} times more efficient than conventional DECP driven with optical pulses, suggesting an intrinsic difference in the force driving the lattice displacement.  

% A similar behavior is observed in our optical time-resolved measurements using 3.10~eV pump and 1.55~eV probe pulses (\hl{see Supplementary Text}). 
% However, a closer inspection of the optical and XUV data under similar nominal fluences shows a much stronger response for the latter (Figure~\ref{fig:fluence}C). Due to the intrinsic complexity in comparing absolute fluences in table-top and FEL experiments, a more compelling quantification of the phonon driving efficiency is obtained by monitoring the coherent response of the system against the transient reflectivity amplitude. In fact, both quantities scale linearly with the pump fluence (\hl{see Figure SX}), providing an internal calibration that is intrinsically independent of the absorbed fluence and nominal photocarrier density. The coherent phonon driving efficiency, here defined as the slope of the linear fit between these two quantities, corresponds to \hl{0.38 and 0.48} for the 3.1~eV and 46.2~eV pump photon energies, respectively. These results confirm the that XDECP in tellurium is \mbox{1.94 $\pm$ 0.08} more efficient than conventional DECP driven with optical light.  

%%%%%%%%%%%%%%%%%%%%%%%%%%%%%%%%%%%%%%%%%%%%%%%%%%%%%%%%
%%%%%%%%%%%%%%%%%%%%%%%%%%%%%%%%%%%%%%%%%%%%%%%%%%%%%%%%
%%%%%%%%%%%%%%%%%%%%%%%%%%%%%%%%%%%%%%%%%%%%%%%%%%%%%%%%
%%%%%%%%%%%%%%%%%%%%%%%%%%%%%%%%%%%%%%%%%%%%%%%%%%%%%%%%

\paragraph{Multiband origin of the core-resonant displacive force:}

The linear scaling of the phonon amplitude with the XUV excitation fluence is consistent with the original formulation of the DECP model \cite{zeiger_theory_1992,merlin1997generating}, which employed tellurium as prototypical system for NIR excitation \cite{cheng_mechanism_1991}. In this framework, an ultrafast excitation across the band gap creates a non-equilibrium carrier distribution that exerts an electrostatic force on the lattice, suddenly displacing its equilibrium position and launching coherent nuclear motion. Because the driving force rises on a timescale shorter than the vibrational period, the nuclei move in phase and modulate the optical properties of the system until decoherence occurs through electron–phonon and phonon–phonon scattering \cite{dhar1994time,pan2025ab}. While this picture successfully describes optical excitations in several materials \cite{zeiger_theory_1992,merlin1997generating,lakehal2019microscopic,caruso_quantum_2023}, its applicability to core-resonant excitation has not been established yet.

In contrast to conventional excitation across the band gap, Te N$_{4,5}$-resonant pumping promotes electrons into the conduction band while leaving an equivalent number of localized 4d$_{3/2}$ and 4d$_{5/2}$ core holes (Figure~\ref{fig:main}C). These holes decay within a few femtoseconds through Auger processes into higher-lying states \cite{auger1923rayons,pollak1972evolution}, thereby defining an initial carrier distribution in which the holes reside in shallow-core levels inaccessible to NIR and visible pulses.
To describe this scenario, we simulate the light-induced dynamics by simultaneously calculating the time evolution of the photocarriers occupation and solving the equation of motion for the lattice \cite{caruso_quantum_2023,emeis2025coherent}. Electron and hole relaxation is treated within the relaxation-time approximation (RTA), yielding a time-dependent electron–phonon driving force $D^{\rm eph}(t)$ acting on the $\mathrm{A_1}$ phonon:
\begin{equation}
  D^{\rm eph}_\mathrm{A_1} (t) = -\frac{2\omega_\mathrm{A_1}}{\hbar} \sum_{n\textbf{k}} g_{nn}^\mathrm{A_1}(\textbf{k},0)\big[f_{n\textbf{k}}(t)-f^{(0)}_{n\textbf{k}}\big],
  \label{eq:Deph}
\end{equation}
where $\omega_\mathrm{A_1}$ is the phonon frequency at zero momentum, 
$g_{nn}^{\mathrm{A_1}}(\textbf{k},0)$ is the EPC matrix element for the \textit{n}-th electronic band with momentum \textbf{k} and at the centre of the phonon dispersion curve, and $f_{n,\textbf{k}}(t)$ is the time-dependent electronic distribution, for band \textit{n} and momentum \textbf{k}, with $f^{(0)}_{n\textbf{k}}$ corresponding to the equilibrium distribution before photoexcitation \cite{emeis2025coherent}.

\begin{figure} 
	\centering
	\includegraphics[width=.7\textwidth]{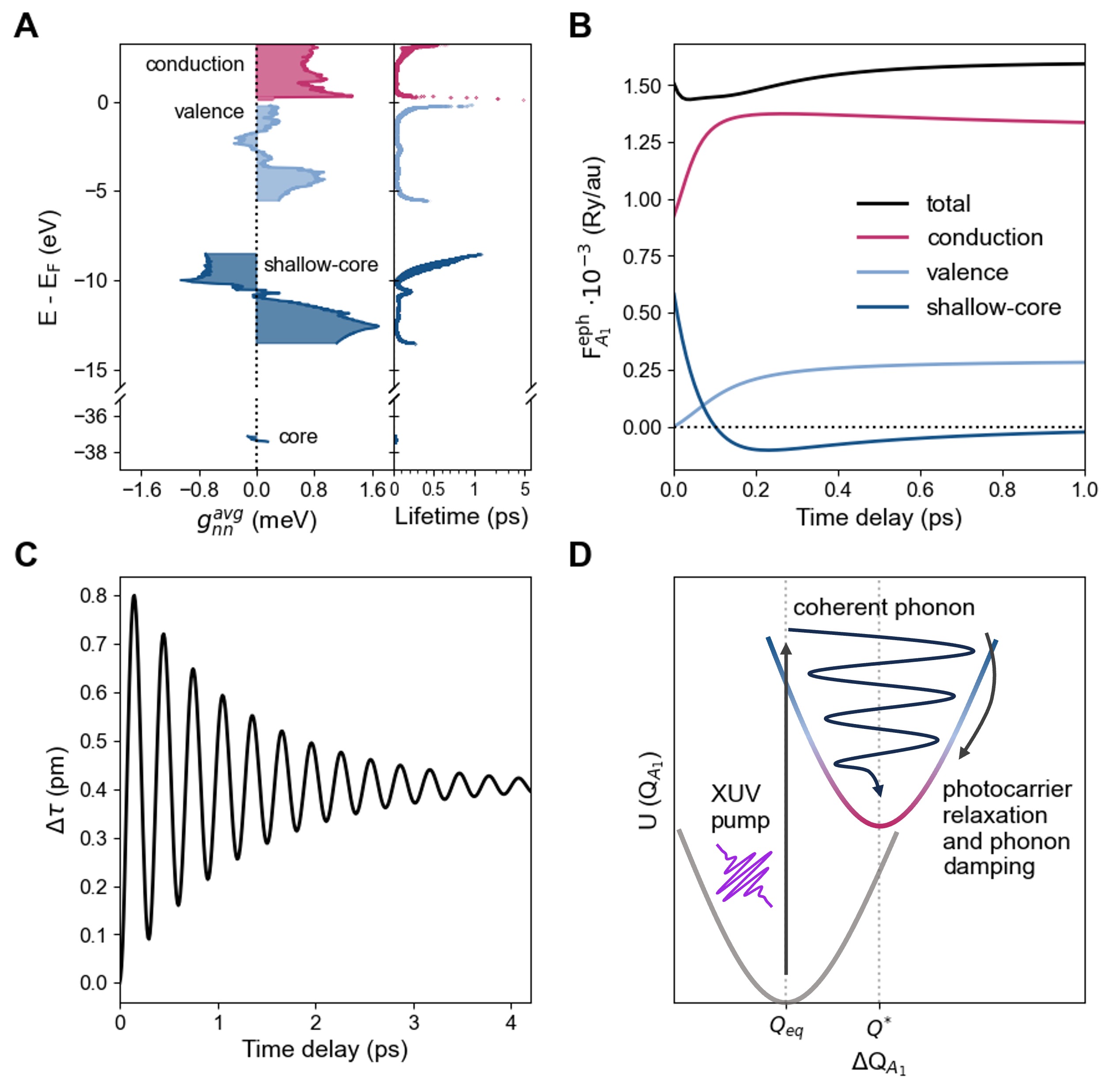}
    \caption{
    \textbf{\textit{Ab initio} calculations of tellurium coherent response upon core-resonant excitation.} 
    (\textbf{A}) Momentum-averaged EPC strength per energy, $g_{nn}^{\mathrm{A_1}, avg}(E)$ (left panel) and room temperature electron lifetimes (right panel) as a function of the relative energy with respect to the Fermi level $\mathrm{E_F}$.    
    (\textbf{B}) Total direct force (black) acting on the atomic sites and relative band contribution (pink: conduction; light-blue: valence; blue: shallow-core) after 46.2~eV excitation with a photocarrier density of $\mathrm{4\cdot10^{20}~{\rm cm}^{-3}}$. 
    (\textbf{C}) Time-dependent atomic site displacement along the $\mathrm{A_1}$ mode upon 46.2~eV excitations.
    % (\textbf{D}) Schematic representation of the core-resonant photodynamics: the XUV excitation induces a sudden shift of the lattice potential $\mathrm{U(Q_{A_1})}$ away from the equilibrium configuration $\mathrm{Q_{eq}}$ towards the photoexcited state minimum centered at the metastable position $Q^{*}$. The EPC force initiates the coherent phonon dynamics, which continues during photocarrier relaxation across the electronic bands, partially lifting the Peierls distortion characterising the ground state of trigonal tellurium.
    % (\textbf{D}) Schematic representation of the X-ray driven photodynamics: the XUV pump induces a sudden shift of the lattice potential $\mathrm{U(Q_{A_1})}$ away from the equilibrium configuration $\mathrm{Q_{eq}}$
    % towards the photoexcited state minimum centered at the metastable position .
    % The EPC force evolves as photocarriers relax across the  band structure, with the contribution of shallow-core holes (blue region of the parabola) progressively compensated by the growth of the conduction band electrons term (pink). The resulting force initiates the coherent phonon dynamics, which continues until complete dephasing.    
    (\textbf{D}) Schematic representation of the FEL driven photodynamics. The XUV pump induces a sudden shift of the lattice potential $\mathrm{U(Q_{A_1})}$ from the equilibrium configuration $Q_{eq}$ towards the photoexcited-state minimum at the metastable position $Q^{*}$, activating the coherent phonon dynamics. As photocarriers relax across the band structure, the EPC force evolves: the shallow-core hole contribution decays and changes sign, while the valence hole and conduction electron terms grow, as indicated by the colour change from blue (shallow-core holes) to light-blue (valence holes) and pink (conduction electrons) of the excited-state potential.
    % following a colour code from blue to pink along the excited-state potential. 
    The phonon oscillation is progressively damped, leaving the lattice displaced towards $Q^{*}$.% until the photocarriers eventually recombine, beyond the simulated time window.    
    }
    \label{fig:theory}
% \vspace{0pt}
\end{figure}

Equation~\eqref{eq:Deph} expresses the displacive force as a sum of band-resolved contributions weighted by both their EPC strengths and transient occupations, which can be obtained using the {\tt EPW} code \cite{lee2023electron} as outlined in the Computational Details. The activation of the phonon is therefore governed jointly by the total density of excited carriers and by their distribution across the band structure, which evolves through scattering.
The momentum-averaged EPC strength per energy, $g_{nn}^{\mathrm{A_1}, avg}(E)$, is shown in the left panel of Figure~\ref{fig:theory}A and reveals that shallow-core and conduction bands provide dominant contributions across the density of state (DOS). 
The sign of the coupling originates from the gradient of the potential energy surface along the $\mathrm{A_1}$ mode, and accounts for the charge of the photocarriers, respectively positive and negative for photocarriers below and above the Fermi level. This sign determines the direction of the force driving the atomic motion, which for the upper shallow-core band is opposite to that of most of the DOS.
% The sign of the coupling accounts for the sign of the charge carriers, respectively positive and negative for photocarriers below and above the Fermi level, and originates from the gradient of the potential energy surface along the $\mathrm{A_1}$ coordinate. 
% It determines the direction of the force driving the atomic motion, which for the upper shallow-core band is opposite to that of most of the DOS.
The right panel of Figure~\ref{fig:theory}A reports the electron lifetimes due to electron-phonon interaction and scattering at room temperature, which influence how long the non-equilibrium occupation lasts and thereby how long these states contribute to the displacive force. Because the 4d core holes decay within \mbox{$\sim \,$1--2}~fs \cite{auger1923rayons,pollak1972evolution} and these orbitals have the smallest EPC (Figure~\ref{fig:theory}A, left), their displacive force is negligible. We therefore model the 46.2~eV excitation by placing electrons in the conduction band, simulating the vertical core-to-conduction band transition,
% through vertical transitions while
and initializing a uniform hole population in the shallow-core levels to account for Auger relaxation. Consistently with the experiment, charge neutrality is preserved within the probed volume \cite{bohinc2019nonlinear}, as the mean free path of photoelectrons generated in tellurium with 46.2~eV photons is sub-nm \cite{nguyen2015penn}, i.e. shorter than the 42~nm penetration depth of the NIR probe \cite{tutihasi_optical_1969}.

The simulations in Figure~\ref{fig:theory}B reveal a pronounced evolution of the different band contributions over sub-picosecond timescales. The shallow-core term (blue curve) has the largest absolute value immediately after excitation but rapidly decreases as holes relax, reversing sign at $\sim \,$100~fs and vanishing within 1~ps.
This change is due to fast population transfer between the shallow-core band peaked at -12.5~eV and the band centered at -9~eV, whose EPC has negative sign (Figure~\ref{fig:theory}A, left panel) and which is characterized by a significantly longer lifetime (Figure~\ref{fig:theory}A, right panel). In contrast, the valence band contribution (light blue curve) grows and saturates over the first 300 fs. Similarly, the conduction band contribution (pink curve) grows as electrons thermalize toward the band minimum where EPC is strongest, ultimately dominating the displacive force. We attribute this behavior to the antibonding character of conduction band orbitals in tellurium \cite{joannopoulos1975electronic}, which weakens covalent bonds when populated \cite{tangney2002density}. 

The competing evolution of electron and hole contributions leads to a gradual buildup of the total direct force toward a plateau value of $\mathrm{\sim \,1.6\cdot10^{-3}~Ry/au}$.
This force drives a coherent lattice dynamics that oscillates at \mbox{$\nu_\mathrm{ph} = 3.31$}~THz, with a maximum amplitude of the atomic displacement $\Delta\tau \sim \,$0.8~pm and a damping time of 1.22~ps (Figure~\ref{fig:theory}C) governed by phonon–phonon scattering \cite{pan2025ab}. 
The net effect is a sudden shift of the lattice potential minimum away from the ground state configuration $\mathrm{Q_{eq}}$, which is stabilized by a Peierls distortion along the $\mathrm{A_1}$ eigenvector \cite{johnson2009full,ning2022light}, and is thus perturbed by the electronic excitation.
% The net effect is a sudden shift of the lattice potential minimum away from the ground state configuration that is initially stabilized by a Peierls distortion along the $\mathrm{A_1}$ eigenvector \cite{johnson2009full,ning2022light}.
% % away from the Peierls-distorted configuration along the $\mathrm{A_1}$ eigenvector that stabilizes the ground state \cite{johnson2009full,ning2022light}. 
A schematic representation of the coupled electronic and nuclear dynamics in the non-equilibrium state is summarized in Figure~\ref{fig:theory}D.

%%%%%%%%%%%%%%%%%%%%%%%%%%%%%%%%%%%%%%%%%%%%%%%%%%%%%%%%
%%%%%%%%%%%%%%%%%%%%%%%%%%%%%%%%%%%%%%%%%%%%%%%%%%%%%%%%
%%%%%%%%%%%%%%%%%%%%%%%%%%%%%%%%%%%%%%%%%%%%%%%%%%%%%%%%
%%%%%%%%%%%%%%%%%%%%%%%%%%%%%%%%%%%%%%%%%%%%%%%%%%%%%%%%

\noindent

\subsection*{Discussion and Conclusion}

Our results experimentally demonstrate the efficient generation of coherent phonons upon core-resonant electronic excitation, establishing the XDECP mechanism.
The measurements are complemented by \textit{ab initio} simulations that include electron–phonon interaction and scattering within the RTA, reproduce the frequency and damping timescales of the observed oscillation (3.31 vs 3.36-3.54~THz and 1.22 vs~1.8 ps, respectively), and decompose the displacive force into separate contributions of electrons and holes occupying distinct regions of the band structure.
% simulations that include electron–phonon interaction and scattering within the RTA, and that reproduce the frequency and damping timescales of the observed oscillation (3.31 vs 3.54~THz and 1.22 vs~1.8 ps, respectively). More importantly, they resolve how the displacive force results from the evolving contributions of electrons and holes distributed across different regions of the band structure.
Due to the computational complexity of the out-of-equilibrium dynamics, our model neglects carrier recombination and secondary carrier multiplication processes, such as impact ionization and intraband Auger relaxation \cite{medvedev2010transient,bohinc2019nonlinear}, thereby maintaining a fixed photocarrier density. Carrier multiplication following XUV excitation has been studied in materials such as silicon \cite{medvedev2010transient}, diamond \cite{medvedev2013nonthermal}, and silicon nitride \cite{bohinc2019nonlinear}, where it leads to a proportional increase of electrons and holes in the conduction and valence bands, respectively (see section \ref{SM:XUV_photocarrier_density} of the SM for an estimation of the photocarrier density). We remark that including these channels would primarily rescale the magnitude of the driving force without altering the mechanism identified here, compatibly with the linear regime of our experiment, and is therefore beyond the scope of the present work.

More critically, the calculation highlights the relevant role of ultrafast carrier relaxation in shaping the electron–phonon force during the first few hundred femtoseconds of dynamics. Within this time window, the displacive force acts as a pulsed trigger, initiating a motion toward a metastable configuration $\mathrm{Q^{*}}$ (Figure~\ref{fig:theory}D) whose response is sensitive to the initially-prepared carrier distribution and subsequent band-specific lifetime (Figure~\ref{fig:theory}A, right). 
Independent evidence for this behavior was recently obtained in Sb, where attosecond XUV transient absorption following NIR excitation resolved a spectral dependence of the $\mathrm{A_{1g}}$ phonon phase and led to the conclusion that the displacive force follows the relaxation of the photoexcited carriers towards the Fermi level rather than their number alone \cite{drescher2025onset}.
% The consequence of hot-carrier thermalization on the onset of the phonon activation in Sb was recently observed in a NIR-driven attosecond transient absorption experiment \cite{drescher2025onset}, further corroborating that tailoring the electron distribution provides a pathway to influence the structural changes. 
Such an effect is expected to be even stronger for core-resonant excitations, and we ascribe the twofold higher driving efficiency of tellurium under XUV excitation to the reshaping of the displacive force, which populates states inaccessible to valence-to-conduction band transitions. 
Additional experimental confirmations would require combining X-ray pumping with probe methods capable of resolving band-selective couplings between photocarriers and optical phonons, such as time-resolved photoemission \cite{gerber2017femtosecond}, or using X-ray diffuse scattering \cite{capotondi2025time,trigo2013fourier,huang2024nanometer,wang2025impulsive} to map how specific electronic states contribute to the activation of distinct phonon branches. 
This concept extends beyond electron-lattice coupling, including exciton-phonon interactions \cite{trovatello2020strongly,krotz2025surface}, state-selective polaron formation \cite{restelli2026ultrafast}, and spin dynamics in antiferromagnets, where displacive excitation of coherent magnons (DECM) has been proposed as the magnetic counterpart of DECP \cite{kalashnikova2008impulsive,lopez2025magneto,jang20234d}. FEL-induced coherent magnons, observed in NiO under circularly polarized core-resonant excitation and not at off-resonant photon energies \cite{simoncig2017generation}, may represent another manifestation of this general principle.

% \hl{[include again comment on EXCITONS: "This strategy extends beyond electron-lattice coupling, including exciton-phonon interactions \cite{trovatello2020strongly} (cite here also the arxiv of Tempelaar) and spin dynamics in antiferromagnets (AFMs)."]}.\\

% The underlying concept extends beyond lattice dynamics. An analogous \hl{impulsive displacement of an order parameter} has been invoked to describe spin dynamics in antiferromagnets, where displacive excitation of coherent magnons (DECM) has been proposed as the magnetic counterpart of DECP \cite{kalashnikova2008impulsive,lopez2025magneto,jang20234d}. FEL-induced coherent magnons observed in NiO under circularly polarized core-resonant excitation, but not at off-resonant photon energies \cite{simoncig2017generation}, may represent another manifestation of this general mechanism.

From a mechanistic viewpoint, core orbitals play an indirect yet critical role: they determine the initial configuration of photoexcited electrons in the conduction band through dipole selection rules and simultaneously enable ultrafast population transfer toward shallow-core states via Auger decay. This process occurs on few-femtosecond timescales \cite{pollak1972evolution}, faster than typical FEL pulse duration or phonon periods, and is intrinsically site-specific because it depends on the spatial overlap of the initial and final state wavefunctions \cite{santra2008concepts}. The large EPC associated with shallow-core states suggests that exciting distinct resonances can modify the balance of band-resolved forces on ultrafast timescales, offering the prospect of element-selective or band-specific lattice activation via X-ray light pulses \cite{ferrer2015nonlinear}. 
In tellurium, the upper shallow-core and conduction bands provide the dominant driving contributions, with opposite sign. Selective photoionization of the former would therefore reverse the sign of the net force with respect to conventional band gap excitation, steering the atomic motion into a chosen structural rearrangement.
Taken together, these results establish XDECP as a mechanism for driving coherent lattice dynamics by core-resonant excitation, and disentangle the carrier contributions that set the magnitude and direction of the displacive force. Because it follows from the resonance condition rather than the photon energy, XDECP extends from the extreme ultraviolet used here to soft and hard X-rays. More broadly, the excited core level itself may become a control parameter for the rational engineering of structural change in condensed matter.
% \hl{In conclusion, the results presented in this work establish core-resonant excitation as an experimentally validated route to drive coherent phonons via XDECP and access band-resolved control parameters for displacive lattice dynamics}.
% Taken together, these results establish XDECP as a mechanism for driving coherent lattice dynamics with X-ray light, and resolve the band-resolved carrier contributions that set the magnitude and direction of the displacive force. 
% More broadly, the excited core level itself may become a control parameter for the rational engineering of structural change in condensed matter.

%%%%%%%%%%%%%%%% MATERIALS AND METHODS %%%%%%%%%%%%%%%

\subsection*{Materials and Methods}

% The Materials and Methods section should contain details of the samples measured,
% experiments performed, observations taken, simulations run, data analysis, statistical methods etc.
% Give enough detail for any competent researcher in your field to fully reproduce the results.

% To refer to this section from the main text, use the numbered note in the reference list \cite{methods}.
% Refer to figures and tables in the same way as in the main text but now all capitalized e.g.
% Figure~\ref{fig:example}, Table~\ref{tab:example},
% Figure~\ref{fig:sup_example} and Table~\ref{tab:sup_example}.
% Cite references in the usual way \cite{example2},
% including any that are only cited in the supplement \cite{sm_example,conference_example}.

% The numbering of figures, tables, equations and pages has been reset to start from S1, as in
% \begin{equation}
% 	\cos(2\theta) = \cos^2\theta - \sin^2\theta.
% 	\label{eq:sup_example} % Use a logical label
% \end{equation}

%%%%%%%%%%%%%%%%%%%%%%%%%%%%%%%%%%%%%%%%%%%%%%%%%%%%%%%%%%%
%%%%%%%%%%%%%%%%%%%%%%%%%%%%%%%%%%%%%%%%%%%%%%%%%%%%%%%%%%%

\subsubsection*{Free-electron laser measurement}

FERMI is a seeded FEL operating in high-gain harmonic generation (HGHG) mode \cite{yu2000high}. The FEL seeding is performed with tunable deep-UV pulses obtained by cascaded frequency up-conversion of a NIR optical parametric amplifier (OPA) pumped by a 50~Hz Ti:Sapphire regenerative amplifier with a central wavelength around 800~nm (1.55~eV) \cite{cinquegrana2021seed}. A NIR pulse generated by a second regenerative amplifier, located on the same optical table and seeded by the same Ti:Sapphire ultrafast oscillator, is optically transported to the EIS-TIMEX end station \cite{masciovecchio_eis_2015} to perform the FEL pump (25~Hz)/NIR probe (50~Hz) measurements with a time jitter of just a few femtoseconds \cite{danailov2014towards}. 
Starting from a seed duration of 65~fs, after the HGHG process the FEL pulses have an estimated duration of about 35~fs \cite{finetti2017pulse}, while the NIR pulse in the range 1.56-1.59~eV is linearly chirped to 0.71~ps by using 10~cm of TBF10 glass. 
The spot sizes on the sample are $170 \times 170$~$\upmu$m$^2$ and $50 \times 50$~$\upmu$m$^2$ full width at half maximum (FWHM) for the pump and probe pulses, respectively. 
We tune the FEL photon energy at 46.2~eV, just above the Te N$_{4,5}$-edge, and employ pulse energies at the sample between 25 nJ and 136 nJ.
No monochromator is used for the FEL pulses, as the spectral broadening factor $\mathrm{\Delta\lambda / \lambda}$ of FERMI emission is between $10^{-3}$ and $10^{-4}$. 
The FEL spectrum is monitored on a shot-to-shot basis using the PRESTO XUV spectrometer \cite{svetina2016presto} located before the experimental chamber. 
In the range of fluences used in our experiment, obtained attenuating the fundamental emission with Al (246~nm) and Mg (316~nm) filters and a tunable N$_2$ gas density, no sample damage is observed. 
The tellurium crystal surface terminates along the (10$\bar{1}$0) plane and is oriented with the \textit{c} axis orthogonal to the wavevector of the FEL and NIR light. 
The pump and probe beams are in quasi-collinear geometry and the sample manipulator is tilted by 10$^{\circ}$ along the in-plane direction in order to reflect the NIR probe into an optical fiber. The fiber output is energy-dispersed using a Kymera 193i spectrometer (Oxford Instruments) equipped with a 1200 line/mm grating and collected on a single-shot basis at 50~Hz using a charge-coupled device (CCD) Andor Newton detector.
The FEL radiation is right-hand circularly polarized, corresponding to the condition of maximal stability of the source, and the probe is linearly polarized along the vertical axis, orthogonal to the \textit{c} axis of the sample. All measurements are performed at room temperature.

%%%%%%%%%%%%%%%%%%%%%%%%%%%%%%%%%%%%%%%%%%%%%%%%%%%%%%%%%%%
%%%%%%%%%%%%%%%%%%%%%%%%%%%%%%%%%%%%%%%%%%%%%%%%%%%%%%%%%%%

\subsubsection*{Computational Details}

We conduct DFT calculations with the plane-wave pseudopotential code {\tt Quantum ESPRESSO}  \cite{giannozzi2009quantum,giannozzi2017advanced},
using a scalar-relativistic projector augmented-wave (PAW) pseudopotential \cite{dal2014pseudopotentials} and the Perdew-Burke-Ernzerhof generalized-gradient approximation in solids (PBEsol) for the exchange-correlation functional \cite{PhysRevLett.100.136406}.
The plane-wave kinetic-energy cutoff is set to 80~Ry, 
and the Brillouin zone is sampled with a $8\times 8 \times 6$ Monkhorst-Pack grid.  
The phonon dispersion is obtained from density-functional perturbation theory (DFPT) on a $4 \times
4 \times 3$ q-point mesh \cite{baroni2001phonons}. 
The initial crystal structure is taken from the Materials Project database \cite{10.1063/1.4812323} and is further optimized. 
The final lattice parameters $a=b=4.6 \ $\r{A} and $c=5.9 \ $\r{A} are reasonably close to the experimental ones \cite{gatti_radial_2020}. 
The electron and phonon eigenvalues, as well as the electron-phonon coupling matrix $g_{nm}^{\rm A_{1}}({\bf k},{\bf q})$, are interpolated on a dense $40\times 40 \times 30$ grid via maximally-localized Wannier functions \cite{marzari2012maximally} within the {\tt EPW} code
\cite{lee2023electron}, which uses {\tt Wannier90} as a library \cite{pizzi2020wannier90}.
Our time-resolved calculations are performed using an \textit{ab initio} 
electron lifetime $\tau^{\rm eph}_{n \bf k}$ (due the electron-phonon interaction),
and the relaxation-time approximation (RTA) for the time evolution of the distribution function: 
    $\partial_t f_{n\bf k}(t) = {[f_{n\mathbf{k}}(t)
- f_{n\mathbf{k}}^{\rm therm}]}/{\tau^{\rm e{-}ph}_{n \bf k}},$
where $\mathbf{k}$ denotes the crystal momentum of an electron in the $n$-th band of the 
thermalised electron distribution $f_{n\mathbf{k}}^{\rm therm}$ (before carrier recombination) for a lattice temperature of $T_\mathrm{lat}=350$~K, and $f_{n\mathbf{k}}(t)$ is the time-dependent electronic occupation. 
%The displacement amplitude of the coherent phonon mode $Q_{\rm A_{1}}$ is obtained by solving the coherent phonon equation of motion $ \partial_t^2 Q_{\rm A_{1}} + \gamma_{\rm A_{1}} \partial_t Q_{\rm A_{1}} + \omega^2_{\rm A_{1}} Q_{ \rm A_{1}} =  - \frac{2\omega_{\rm A_{1}}}{\hbar} \sum_{n \mathbf{k}} g_{nn}^{\rm A_{1}} (\mathbf{k},0)  \Delta f_{n\mathbf{k}}(t)$, which is propagated using the 2nd order Runge-Kutta method. 
%In this equation of motion, $\omega_{\rm A_{1}}$ is the frequency of the coherent phonon mode and $\Delta f_{n\mathbf{k}}(t) = f_{n\mathbf{k}}(t) - f_{n\mathbf{k}}^{\rm (0)}$ is the change of the electronic occupation compared to the carrier distribution before photoexcitation $f_{n\mathbf{k}}^{\rm (0)}$. 
The displacement amplitude of the coherent phonon mode $Q_{\rm A_{1}}$ is obtained by solving the coherent phonon equation of motion $ \partial_t^2 Q_{\rm A_{1}} + \gamma_{\rm A_{1}} \partial_t Q_{\rm A_{1}} + \omega^2_{\rm A_{1}} Q_{ \rm A_{1}} =  D^{\rm eph}_\mathrm{A_1} (t)$, which is propagated using the 2nd order Runge-Kutta method. 
In this equation of motion, $\gamma_{\rm A_{1}}$ is the coherent phonon damping rate, $\omega_{\rm A_{1}}$ is the frequency of the coherent phonon mode as determined by DFPT phonon calculations, and $D^{\rm eph}_\mathrm{A_1} (t)$ is the time-dependent electron–phonon driving force as given in Eq.~\eqref{eq:Deph} for a photocarrier density of $\mathrm{4\cdot10^{20}~{\rm cm}^{-3}}$.
Further details on the theory and implementation can be found elsewhere \cite{emeis2025coherent}.
The coherent phonon damping rate $\gamma_{\rm A_{1}}$ was computed using the method described in Ref.\cite{pan2025ab}.
The third-order force constant, needed for this method, is obtained from a DFT $3 \times 3 \times 2$ supercell calculation as implemented in the {\tt third-order.py} utility of {\tt ShengBTE} \cite{ShengBTE_2014}.

The direct force $F_{\kappa\alpha}(t)$ acting on each atom of the unit cell was computed by substituting the atomic displacement:

\begin{equation}
\Delta \tau_{\kappa\alpha} = \sqrt{\frac{\hbar}{2M_\kappa \omega_{A_1}}} e^{\kappa\alpha}_{A_1} Q_{A_1}
  \label{eq:Q_to_tau}
\end{equation}

into Newton's equation of motion, obtaining:

\begin{equation}
F_{\kappa\alpha}(t) = e^{\kappa\alpha}_{A_1} \sqrt{\frac{M_\kappa\hbar}{2\omega_{A_1}}} D^{\mathrm{eph}}_{A_1}(t)
  \label{eq:D_to_F}
\end{equation}

Here, $\kappa$ is the index of the atoms, $\alpha$ denotes the Cartesian direction, $e^{\kappa\alpha}_{A_1}$ is the phonon eigenvector, $\omega_{A_1}$ is the phonon frequency, and $M_\kappa$ is the nuclear mass.

%%%%%%%%%%%%%%%% MAIN TEXT FIGURES %%%%%%%%%%%%%%%

% Figure1: main results
\begin{figure} 
	\centering
	\includegraphics[width=.9\textwidth]{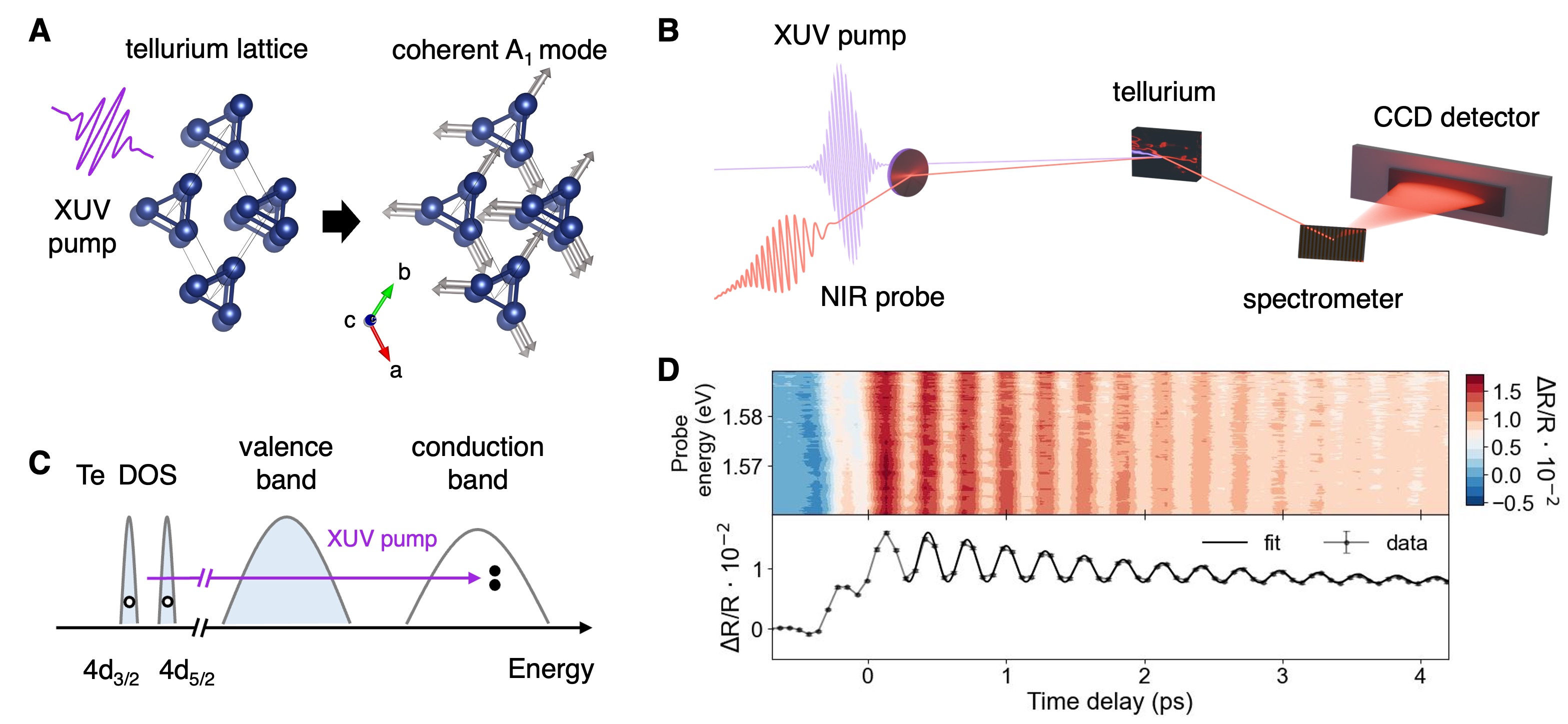}
    \caption{
    \textbf{Displacive activation of tellurium $\mathbf{A_1}$ coherent phonon mode through core-resonant XUV excitation}.
    (\textbf{A}) An ultrashort FEL pulse is tuned at the Te N$_{4,5}$-edge, driving coherent nuclear oscillations along a totally symmetric $\mathrm{A_1}$ phonon mode. 
    The blue spheres represent Te atoms and the gray arrows show the lattice displacements along the normal mode. 
    (\textbf{B}) Scheme of the experimental setup.
    A 35~fs XUV pump photoexcites the tellurium crystal, whose response is monitored as a function of time using a delayed chirped NIR probe pulse. 
    The reflected optical signal is energy dispersed with a grating and collected on a shot-to-shot basis with a charge-coupled device (CCD) camera. 
    (\textbf{C}) Schematic representation of the tellurium density of states (DOS) and XUV-driven electronic transitions at the Te N$_{4,5}$-edge. 
    (\textbf{D}) Chirp-corrected transient reflectivity map for core-resonant excitation using FEL pulses centered at 46.2~eV (0.09~mJ$\cdot$cm$^{-2}$ incident fluence). The lower panel shows the spectrally averaged time trace (connected dots) and the corresponding fit with Eq.~\eqref{eq:transients}. The $\mathrm{A_1}$ coherent oscillation at 3.535~THz dominates the transient signal, persisting beyond 4~ps. 
    %\hl{The error bars are the 95\% confidence interval standard error of the averaged trace}.
    Error bars are $\pm 2\sigma$ (standard error, $\sim$95\% confidence interval) of the pixel-averaged trace, corrected for the measured pixel-to-pixel correlation length of 8.6 pixels.
    }
    
    \label{fig:main}
\end{figure}

% Figure2: fluence dependence
\begin{figure} 
	\centering
	\includegraphics[width=.99\textwidth]{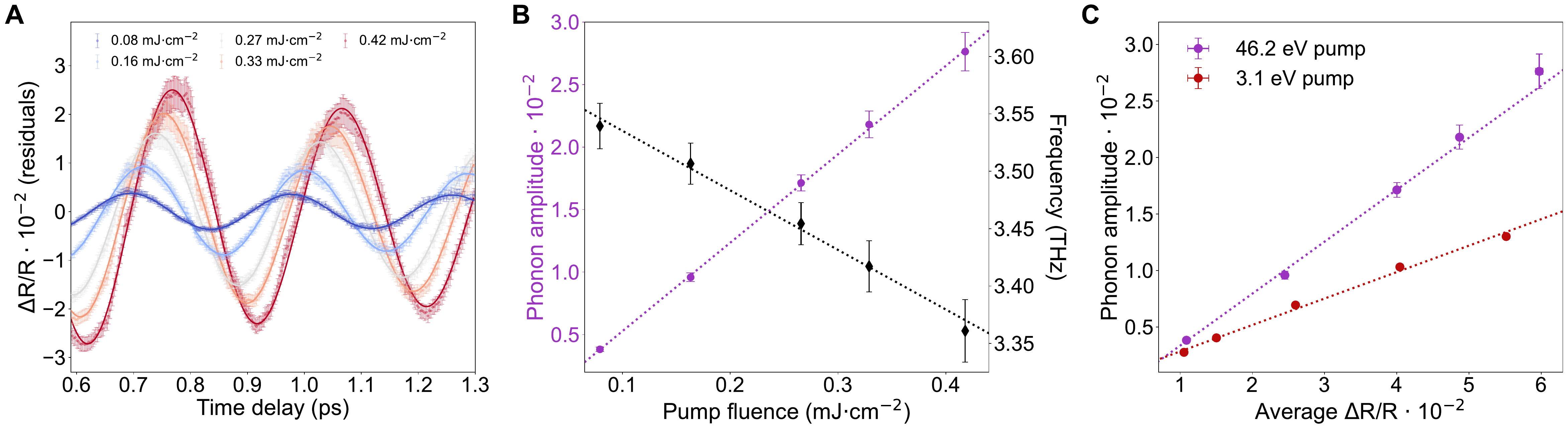}
    \caption{
    \textbf{Efficiency of the X-ray driven displacive excitation of coherent phonons in tellurium.} 
    (\textbf{A}) Coherent phonon response of tellurium (dots) as a function of the FEL pump fluence at 46.2~eV and their best fits (full lines). The incoherent signal from equation~\eqref{eq:transients} is subtracted from the data.
    (\textbf{B}) Phonon amplitude (purple dots, left axis) and frequency (black diamonds, right axis) as a function of the XUV pump fluence, together with their linear fits (dotted lines).
    (\textbf{C}) Phonon amplitude as a function of the average transient reflectivity signal for the core-resonant pump at 46.2 eV (purple) and optical pump at 3.10 eV (red), and their linear fits (dotted lines). The slopes correspond to the phonon driving efficiency for the two excitation photon energies.
    In all panels, error bars are $\pm 2\sigma$ ($\sim$95\% confidence interval). In (\textbf{A}) $\sigma$ is the total uncertainty of the transient reflectivity for each pixel. In (\textbf{B}) and (\textbf{C}) the standard deviation of the fitted parameters is computed with a bootstrap method.
    }
    \label{fig:fluence}
\end{figure}

% Figure3: theory

%%%%%%%%%%%%%%%% REFERENCES %%%%%%%%%%%%%%%

\clearpage 

\bibliography{tellurium_bibliography} 
\bibliographystyle{sciencemag}

%%%%%%%%%%%%%%%% ACKNOWLEDGEMENTS %%%%%%%%%%%%%%%

\newpage

\section*{Acknowledgments}

\paragraph*{Funding:}

O.C. gratefully acknowledges the BNF program (University of Bern) under the project 5320, the Swiss National Science Foundation Postdoc mobility program under the grant agreement P500PN\_214151, and the European Union’s Horizon Europe research and innovation programme under the Marie Sk\l odowska-Curie METRICS HORIZON-MSCA-2022-PF-EF grant agreement no. 101106352. F.Cal. and O.C. acknowledge funding from Cluster of Excellence ‘CUI: Advanced Imaging of Matter’ of the Deutsche Forschungsgemeinschaft (DFG)—EXC 2056—project ID 390715994.
O.D. and G.C. acknowledge financial support from European Union's NextGenerationEU Investment 1.1, PRIN 2022 PNRR HAPPY (ID P20224AWLB, CUP D53D23016720001).
R.Co. gratefully acknowledges financial support from European Union's NextGenerationEU M4C2 - Investment 1.3, PNRR MUR project No. PE0000023-NQSTI.
F.Caru. and C.E. gratefully acknowledge funding by the Deutsche Forschungsgemeinschaft (DFG), Project No. 499426961 and computing time on the high-performance computer Lichtenberg II at TU Darmstadt, funded by the German Federal Ministry of Education and Research (BMBF), and the State of Hesse Ministry of Science and Research, Art and Culture (HMWK), Project No. p0021280.
This work was funded by the European Union under HE-GA 101131771 Lasers4EU 23101, 25998 and 30116. Views and opinions expressed are however those of the author(s) only and do not necessarily reflect those of the European Union. The European Union cannot be held responsible for them.
We acknowledge Elettra-Sincrotrone Trieste for providing access to its synchrotron radiation facilities and we thank Stefano Nannarone for his support during the synchrotron measurements at the BEAR beamline. The authors acknowledge Felix Ritzkowsky for the rendering in Figure~\ref{fig:main}B. 

% This work was supported by the European Union's Horizon 2020 research and innovation programme under \hl{grant agreement no. 23101, 25998 and 30116 Laserlab-Europe}. 

%%%%%%%%%%%%%%%%%%%%%%%%%%%%%%%%%%%%%%%%%%%%%%%%%%%%%%%%%%%
%%%%%%%%%%%%%%%%%%%%%%%%%%%%%%%%%%%%%%%%%%%%%%%%%%%%%%%%%%%

\paragraph*{Author contributions:}
Conceptualization: O.C.; Methodology: M.P. and O.C.; FEL investigation: O.D., R.Cl., G.B., G.F., R.Co., A.A., S.R., P.C., R.G.C., D.D.A., E.Pa., Z.E., M.K., G.K., M.D., L.G., R.M., E.Pr., A.C., G.D.N., C.M., L.F., T.S., F. Carb., F. Cal., M.P., O.C.; Synchrotron measurements: O.D., R.Co., A.G., S.N., M.P., O.C.; Optical investigation: O.D., M.M., A.C., G.C., F.Carb., M.P., O.C.; Chirp detection design: O.D., G.B., G.F., G.K., M.D., L.F., T.S., M.P., O.C.; \textit{Ab initio} calculations: C.E., F.Caru.; Data analysis: O.D., R.Cl., C.E., P.C., M.P., O.C.; Writing-original draft: O.D., R.Cl., C.E., F.Caru., M.P., O.C.; Writing-review and editing: all authors.

%%%%%%%%%%%%%%%%%%%%%%%%%%%%%%%%%%%%%%%%%%%%%%%%%%%%%%%%%%%
%%%%%%%%%%%%%%%%%%%%%%%%%%%%%%%%%%%%%%%%%%%%%%%%%%%%%%%%%%%

\paragraph*{Competing interests:}

There are no competing interests to declare.

%%%%%%%%%%%%%%%%%%%%%%%%%%%%%%%%%%%%%%%%%%%%%%%%%%%%%%%%%%%
%%%%%%%%%%%%%%%%%%%%%%%%%%%%%%%%%%%%%%%%%%%%%%%%%%%%%%%%%%%

\paragraph*{Data and materials availability:}

The data that support the findings of this article will be openly available upon publication.

%%%%%%%%%%%%%%%% SUPPLEMENT LIST %%%%%%%%%%%%%%%

\subsection*{Supplementary materials}
Supplementary Text\\
Figures S1 to S4\\
References \textit{(85-\arabic{enumiv})}\\ % automatically fills out the last reference number

%%%%%%%%%%%%%%%% END OF MAIN TEXT %%%%%%%%%%%%%%%

\newpage

%%%%%%%%%%%%%%%%%%%%%%%%%%%%%%%%%%%%%%%%%%%%%%%%%%%%%%%%%%%
%%%%%%%%%%%%%%%%%%%%%%%%%%%%%%%%%%%%%%%%%%%%%%%%%%%%%%%%%%%
%%%%%%%%%%%%%%%%%%%%%%%%%%%%%%%%%%%%%%%%%%%%%%%%%%%%%%%%%%%
%%%%%%%%%%%%%%%%%%%%%%%%%%%%%%%%%%%%%%%%%%%%%%%%%%%%%%%%%%%
%%%%%%%%%%%%%%%%%%%%%%%%%%%%%%%%%%%%%%%%%%%%%%%%%%%%%%%%%%%
%%%%%%%%%%%%%%%%%%%%%%%%%%%%%%%%%%%%%%%%%%%%%%%%%%%%%%%%%%%
%%%%%%%%%%%%%%%%%%%%%%%%%%%%%%%%%%%%%%%%%%%%%%%%%%%%%%%%%%%
%%%%%%%%%%%%%%%%%%%%%%%%%%%%%%%%%%%%%%%%%%%%%%%%%%%%%%%%%%%
%%%%%%%%%%%%%%%%%%%%%%%%%%%%%%%%%%%%%%%%%%%%%%%%%%%%%%%%%%%
%%%%%%%%%%%%%%%%%%%%%%%%%%%%%%%%%%%%%%%%%%%%%%%%%%%%%%%%%%%
%%%%%%%%%%%%%%%%%%%%%%%%%%%%%%%%%%%%%%%%%%%%%%%%%%%%%%%%%%%
%%%%%%%%%%%%%%%%%%%%%%%%%%%%%%%%%%%%%%%%%%%%%%%%%%%%%%%%%%%
%%%%%%%%%%%%%%%%%%%%%%%%%%%%%%%%%%%%%%%%%%%%%%%%%%%%%%%%%%%
%%%%%%%%%%%%%%%%%%%%%%%%%%%%%%%%%%%%%%%%%%%%%%%%%%%%%%%%%%%

% %%%%%%%%%%%%%%%% START OF SUPPLEMENT %%%%%%%%%%%%%%%

\clearpage   % NOT \newpage — flushes deferred main-text floats

%%%%%%%%%%%%%%%% START OF SUPPLEMENT %%%%%%%%%%%%%%%
\setcounter{figure}{0}
\setcounter{table}{0}
\setcounter{equation}{0}
\setcounter{section}{0}
\setcounter{page}{1}

\renewcommand{\thefigure}{S\arabic{figure}}
\renewcommand{\thetable}{S\arabic{table}}
\renewcommand{\theequation}{S\arabic{equation}}
\renewcommand{\thesection}{S\arabic{section}}
\renewcommand{\thepage}{S\arabic{page}}

% hyperref anchor names — without these, resetting counters creates
% duplicate PDF destinations and links jump to the main-text object
\renewcommand{\theHfigure}{S\arabic{figure}}
\renewcommand{\theHtable}{S\arabic{table}}
\renewcommand{\theHequation}{S\arabic{equation}}
\renewcommand{\theHsection}{S\arabic{section}}
\def\theHpage{S\arabic{page}} % it was: \renewcommand{\theHpage}{S\arabic{page}}

%%%%%%%%%%%%%%%% SUPPLEMENT TITLE PAGE %%%%%%%%%%%%%%%

\begin{center}
\section*{Supplementary Materials for\\ \scititle}

\author{
    Oleg~Dogadov$^{1,2}$,
    Remi~Claude$^{3}$,
    Christoph~Emeis$^{4}$,
    Giovanni~Batignani$^{5,6}$,
    Giuseppe~Fumero$^{5\dagger}$,
    Roberto~Costantini$^{7,8}$,
    Agata~Azzolin$^{9,10,11}$,
    Sabine~Rockenstein$^{9,10,12}$,
    Paolo~Cattaneo$^{3}$,
    Rebeca Gomez~Castillo$^{13}$,
    Matteo~Manzi$^{13}$,
    Angelo~Giglia$^{7}$,
    % \hl{Stefano~Nannarone}$^{7}$,
    Dario De~Angelis$^{14}$,
    Ettore~Paltanin$^{14}$,
    Zeinab~Ebrahimpour$^{14}$,
    Marija~Krstulovic$^{14}$,
    Gabor~Kurdi$^{14}$,
    Miltcho~Danailov$^{14}$,
    Luca~Giannessi$^{14,15}$,
    Riccardo~Mincigrucci$^{14}$,
    Emiliano~Principi$^{14}$,
    Alberto~Crepaldi$^{1}$,
    Giulio~Cerullo$^{1,16}$,
    Giovanni De~Ninno$^{14,17}$,
    Claudio~Masciovecchio$^{14}$,
    Laura~Foglia$^{14}$,
    Tullio~Scopigno$^{5,6,18}$,
    Fabrizio~Carbone$^{3,13}$,
    Francesca~Calegari$^{9,10,11}$, \and
    Fabio~Caruso$^{4}$,
    Michele~Puppin$^{13}$,
    Oliviero~Cannelli$^{9,11,13*}$

	\small$^{1}$Dipartimento di Fisica, Politecnico di Milano, Piazza Leonardo da Vinci, 32, 20133 Milano, Italy.\\
	\small$^{2}$Department of Physical Chemistry, Fritz Haber Institute of the Max Planck Society, 14195 Berlin, Germany.\\
	\small$^{3}$Laboratory for Ultrafast Microscopy and Electron Scattering (LUMES), Institute of Physics, École Polytechnique Fédérale de Lausanne (EPFL), Lausanne 1015 CH, Switzerland.\\
    \small$^{4}$Institute of Theoretical Physics and Astrophysics, Kiel University, 24118 Kiel, Germany.\\
    \small$^{5}$Dipartimento di Fisica, Sapienza Università di Roma, Roma, Italy.\\
    \small$^{6}$Istituto Italiano di Tecnologia (IIT), Center for Life Nano Science @Sapienza, Roma, Italy.\\
    \small$^{7}$CNR—Istituto Officina dei Materiali (IOM), S.S. 14 km 163.5, 34149 Trieste, Italy.\\
    \small$^{8}$Dipartimento di Fisica, Università di Trieste, Via Valerio 2, I-34127 Trieste, Italy.\\
    \small$^{9}$Center for Free-Electron Laser Science, DESY, Notkestraße 85, 22607 Hamburg, Germany.\\
    \small$^{10}$Physics Department, University of Hamburg, Luruper Chausee 149, 22761 Hamburg, Germany.\\
    \small$^{11}$The Hamburg Centre for Ultrafast Imaging, University of Hamburg, Luruper Chausee 149, 22761 Hamburg, Germany.\\
    \small$^{12}$Max-Planck-Institut für Struktur und Dynamik der Materie, Hamburg, Germany.\\
    \small$^{13}$Lausanne Centre for Ultrafast Science (LACUS), École Polytechnique Fédérale de Lausanne (EPFL), CH-1015 Lausanne, Switzerland.\\
    \small$^{14}$Elettra-Sincrotrone Trieste S.C.p.A., Strada Statale 14, km 163.5, AREA Science Park, I-34149, Basovizza, Trieste, Italy.\\
    \small$^{15}$Istituto Nazionale di Fisica Nucleare, Laboratori Nazionali di Frascati, 00044 Frascati, Italy.\\
    \small$^{16}$CNR-IFN, Piazza Leonardo da Vinci 32, 20133 Milan, Italy.\\
    \small$^{17}$Laboratory of Quantum Optics, University of Nova Gorica, Si-5270 Ajdovščina, Slovenija.\\
    \small$^{18}$Istituto Italiano di Tecnologia (IIT), Graphene Labs, Genova, Italy.\\
    \small$^\dagger$Present address: CNR NANOTEC, Institute of Nanotechnology, Via Monteroni, Lecce 73100, Italy.\\
	\small$^\ast$Corresponding author. Email: oliviero.cannelli@cfel.de
}

\end{center}

% Fill out the numbers for each type of supplementary material,
% and delete any lines that aren't applicable.
% These are just example numbers that don't match the rest of this template.
\subsubsection*{This PDF file includes:}
Sections S1 to S5\\
Figures S1 to S4\\

\newpage

%%%%%%%%%%%%%%%%%%%%%%%%%%%%%%%%%%%%%%%%%%%%%%%%%%%%%%%%%%%
%%%%%%%%%%%%%%%%%%%%%%%%%%%%%%%%%%%%%%%%%%%%%%%%%%%%%%%%%%%

\section{Synchrotron measurement}\label{SM:synchrotron}
\begin{figure}[p]%{0.5\textwidth}
\vspace{-0pt}\centering
    \includegraphics[width=.45\linewidth]{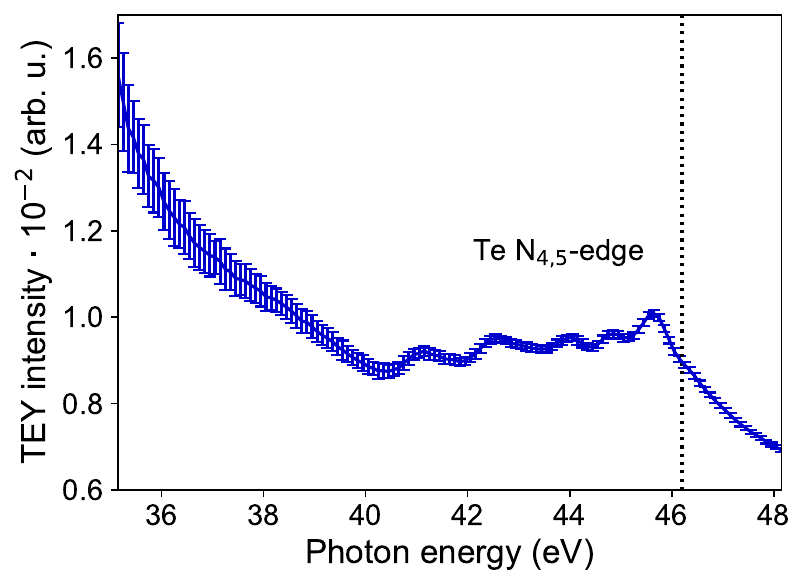}
    \caption{
    \textbf{X-ray absorption spectrum of trigonal tellurium in the XUV range}. The dotted vertical line indicates the FEL photon energy used in the experiment (46.2~eV), tuned in resonance just above the Te N$_{4\text{,}5}$-edge. The spectrum is collected in total electron yield mode at room temperature. The error bars denote the $\pm 2\sigma$ (standard error, $\sim$95\% confidence interval) of the measurement.
    }
    \label{fig:XAS}
% \vspace{0pt}
\end{figure}

The X-ray absorption spectroscopy (XAS) spectrum of trigonal tellurium is collected at the BEAR beamline of the Elettra synchrotron in Trieste \cite{nannarone2004bear}. We measure the same sample used in the FEL experiment, \textit{i.e.}, the crystal with surface termination along the (10$\bar{1}$0) plane, oriented with the \textit{c} axis orthogonal to the synchrotron beam polarization and wavevector. The crystal is mounted on a metallic base plate using conductive silver paste. The synchrotron beam is monochromatized using the normal incidence monochromator based on a plane mirror working at 15$^{\circ}$ and a plane laminar grating with 1200~lines/mm to scan the energy range 35.15--48.15~eV, and is focused using an ellipsoidal mirror. The spot size on the sample is reduced to $130 \times 300 \ \upmu$m${}^2$ by using horizontal and vertical slits. Higher order harmonics are suppressed using a Mg filter.
The tellurium signal is collected in total electron yield (TEY) mode. 
Synchrotron flux fluctuations are compensated by normalizing the TEY signal by the photocurrent measured on the ellipsoidal mirror placed immediately before the sample.
A transmission spectrum is collected in the same energy range using a silicon photodiode in absence of the sample, scaled by the photocurrent measured with the same ellipsoidal refocusing mirror placed before the sample.
In all cases, a Keithley 6517A ammeter is used. 
Prior to normalization, a dark current signal measured at the same current level of each detector is subtracted from the corresponding detector.  
The XAS spectrum of tellurium is obtained scaling the TEY spectrum by the photodiode spectrum, thereby removing any energy-dependent contribution coming from the beamline optical elements. The energy axis is calibrated measuring the transmission signal of a Mg filter. The obtained spectrum is reported in Figure~\ref{fig:XAS}.

%%%%%%%%%%%%%%%%%%%%%%%%%%%%%%%%%%%%%%%%%%%%%%%%%%%%%%%%%%%
%%%%%%%%%%%%%%%%%%%%%%%%%%%%%%%%%%%%%%%%%%%%%%%%%%%%%%%%%%%

\section{XUV transient reflectivity analysis} \label{SM:XUV_analysis}

\subsection{Transient reflectivity map}\label{SM:XUV_TRR}
XUV transient reflectivity data is recorded using chirped NIR pulses obtained by inserting 10~cm of TBF10 glass in the probe arm. The chirp is essentially linear over the entire probe photon energy range. However, due to the non-flat spectral phase of the probe, a fourth order polynomial function is required to correctly describe the dispersion curve.
The transient reflectivity map is collected by scanning the relative delay between the XUV pump and the NIR probe. At each time delay, 500 pump-on/pump-off pairs are recorded on a shot-to-shot basis at 50 Hz (25 Hz FEL chopping rate). The pulses are filtered based on their average FEL-pump intensity and NIR-probe intensity, keeping only the pulses that are within twice the standard deviation from the average value, resulting in approximately 450 shot pairs per delay.
A detector background, taken as the mean over 100 pixels where no probe signal is present, is subtracted from each acquisition before averaging the pump-on and pump-off pulses. For each delay \textit{t}, the transient reflectivity $\Delta R(t) / R$ is defined as $\mathrm{(R_\mathrm{\, FEL_{on}}/R_\mathrm{\, FEL_{off}})-1}$, thereby removing the laser correlation on short timescales, and the corresponding standard error is computed.
The map is then dechirped by interpolating each pixel onto a common delay axis. The mean trace is obtained by averaging the signal in the 1.56-1.59~eV range and its uncertainty is computed as standard error, scaled to account for a statistical pixel-to-pixel correlation length of 8.6 pixels.

\subsection{Fluence dependence traces}\label{SM:XUV_fluences}
For the fluence dependence scans, every transient spectrum is the average of 29 independent acquisitions. Each acquisition follows the same analysis of the transient reflectivity map at a single time delay as detailed in section \ref{SM:XUV_TRR}: collection of 500 pump-on/pump-off shot pairs, filtering, averaging, background subtraction, computation of the transient reflectivity signal as a function of the probe photon energy. The statistical uncertainty at every detector pixel is the standard error of the mean of the 29 acquisitions. However, this value underestimates the experimental error, which is dominated by a systematic component observed to be proportional to the transient signal. Therefore, the total error is computed by adding in quadrature the statistical error and a multiplicative term: $\mathrm{\sigma^2 = \sigma^2_{stat} + (k \cdot |\Delta R/R|)^2}$. A single value of $\mathrm{k=0.0187}$ is used for the whole fluence series, determined by requiring the reduced $\chi^2$ over all fluences to be equal to 1 when the traces are fitted with equation~\eqref{eq:transients} of the main manuscript; after including this term, the individual reduced $\chi^2$ lie between 0.7 and 1.5, against inital values of 2.3-32.

\begin{figure}[p]%{0.5\textwidth}
    \centering
    \includegraphics[width=.9\linewidth]{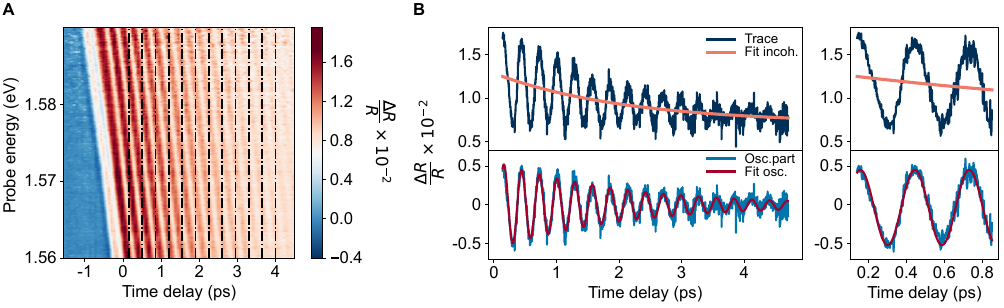}
    \caption{
    \textbf{Photon-energy to time-delay conversion of chirped transient reflectivity signal upon core-resonant XUV excitation.} (\textbf{A}) Transient reflectivity after excitation at 46.2~eV prior to chirp correction (chirp-corrected data are shown in the main text, Figure~\ref{fig:main}D). 
    (\textbf{B}) Left: Time trace obtained from photon-energy to time-delay conversion of several transient spectra collected at equally spaced time delays, marked with black vertical lines in panel (\textbf{A}). 
    Right: Single transient spectrum after photon-energy to time-delay conversion.
    }
    \label{fig:crocodile}
% \vspace{0pt}
\end{figure}

In the fluence dependence series, the time traces are computed converting the probe photon energy in time delays following the same strategy of the chirped impulsive stimulated Raman scattering scheme developed in the optical domain \cite{batignani_broadband_2019}. 
Within this approach, the spectra are recorded at a fixed time delay between pump and probe beams, but the dynamics is monitored by the individual spectral components of the probe, which interact with the sample at different delays due to the temporal chirp introduced by the TBF10 glass.
The starting time delay of 0.59~ps, defined as the arrival time of the probe component at 1.59~eV with respect to the XUV pump, is determined by comparing the traces with photon energy-domain signal of the transient reflectivity map at the same time delay.
The photon-energy to time-delay calibration curve is obtained through a direct analysis of the chirped transient reflectivity scan (Figure~\ref{fig:crocodile}A). This map reports the same transient reflectivity data shown in Figure~\ref{fig:main}D of the main text, but before chirp correction.
The delay of each probe photon energy is computed by tracking the arrival time of the third minimum of the oscillatory component of the transient signal. This reference is chosen to avoid possible contributions from cross-phase modulation close to the zero delay of each pixel.
The validity of the calibration curve is confirmed through the analysis of the dynamics after conversion. The precision of our results is enhanced by combining multiple energy traces (vertical black lines in Figure~\ref{fig:crocodile}A) into a single curve spanning 4~ps (Figure~\ref{fig:crocodile}B). 
The incoherent and coherent fits using equation~\eqref{eq:transients} of the main text are separately reported in the top and bottom panels of Figure~\ref{fig:crocodile}B (left).
We obtain \mbox{$\tau_\mathrm{el} = 2.1 \pm 0.3$}~ps, \mbox{$\tau_{damp} = 2.0 \pm 0.1$}~ps and \mbox{$\nu_\mathrm{ph} = 3.53 \pm 0.01$}~THz, which are in excellent agreement with the time-domain results. 
We note that the analysis of the single photon energy traces spanning 0.71~ps (Figure~\ref{fig:crocodile}B, right panels) provides a phonon frequency of $3.48 \pm 0.07$~THz, which also agrees, within the error bars, with both time-domain and photon-energy to time-delay converted data. The error bars of the fitting parameters in Figure~\ref{fig:fluence} of the main text are obtained by a moving-block bootstrap method (400 replicas, $\mathrm{\tau_{el}}$ set to the transient reflectivity map value) and computing the standard deviation of the resulting parameter distribution.

%%%%%%%%%%%%%%%%%%%%%%%%%%%%%%%%%%%%%%%%%%%%%%%%%%%%%%%%%%%
%%%%%%%%%%%%%%%%%%%%%%%%%%%%%%%%%%%%%%%%%%%%%%%%%%%%%%%%%%%
\section{Optical transient reflectivity analysis}

\subsection{Transient reflectivity map}\label{SM:TRR_map}
\begin{figure}[p]%{0.5\textwidth}
    \centering
    \includegraphics[width=.9\linewidth]{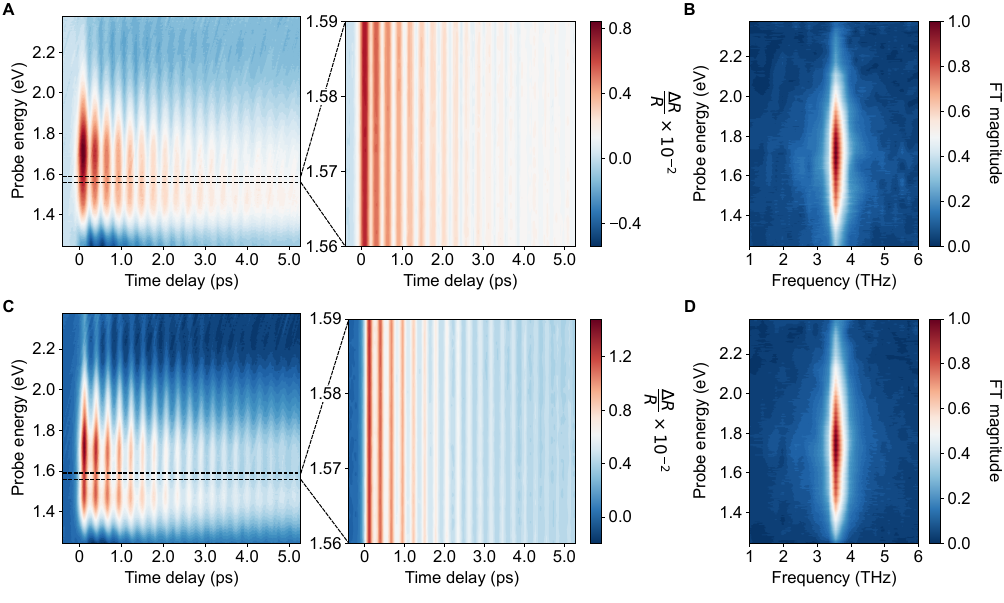}
    \caption{
    \textbf{Optical transient reflectivity of tellurium.} (\textbf{A}) Transient reflectivity map (left) after 3.10~eV pump, 0.58~mJ/cm${}^2$, probe polarization parallel to the $c$~axis, and a zoom on the 1.56--1.59~eV region (right). 
    (\textbf{B}) Fourier transform (FT) of the oscillatory component of the signal in \textbf{(A)} showing a single mode at 3.60~THz.
    (\textbf{C-D}) Same as \textbf{(A--B)} for the probe polarization orthogonal to the $c$~axis.
    }
    \label{fig:opt_pp}
% \vspace{0pt}
\end{figure}

Optical transient reflectivity measurements is performed covering the spectral range used in the XUV-pump--NIR-probe experiment. In these all-optical measurements, the sample is excited with the second harmonic of Ti:Sapphire laser (Coherent, Libra) at 3.1~eV, 0.58~mJ/cm${}^2$ pump fluence. %in the pump fluence range 0.41-2.00~mJ/cm${}^2$.
The photoinduced dynamics is tracked with broadband probe pulses, generated in a YAG plate, pumped with the 1.0~eV signal of a home-built optical parametric amplifier. 
Similarly to the XUV excitation, the pump is circularly polarized, whereas the probe is linearly polarized either parallel or orthogonal to the \textit{c} axis of the sample.
The measurements are performed at room temperature and 1~kHz repetition rate. 
The results are summarized in Figure~\ref{fig:opt_pp}, which shows a uniform transient signal in the energy region of interest 1.56--1.59~eV. No narrow features are observed, as expected from steady-state optical properties of tellurium \cite{tutihasi_optical_1969}.
In the whole probe range, we observe only one phonon mode at 3.60~THz, as reported in Figure~\ref{fig:opt_pp}B,D. 

\section{Comparison of the XUV and optical transient reflectivity as a function of the pump fluence}\label{SM:comparison_XUV_opt_fluences}
\begin{figure}[p]%{0.5\textwidth}
    \centering
    \includegraphics[width=.9\linewidth]{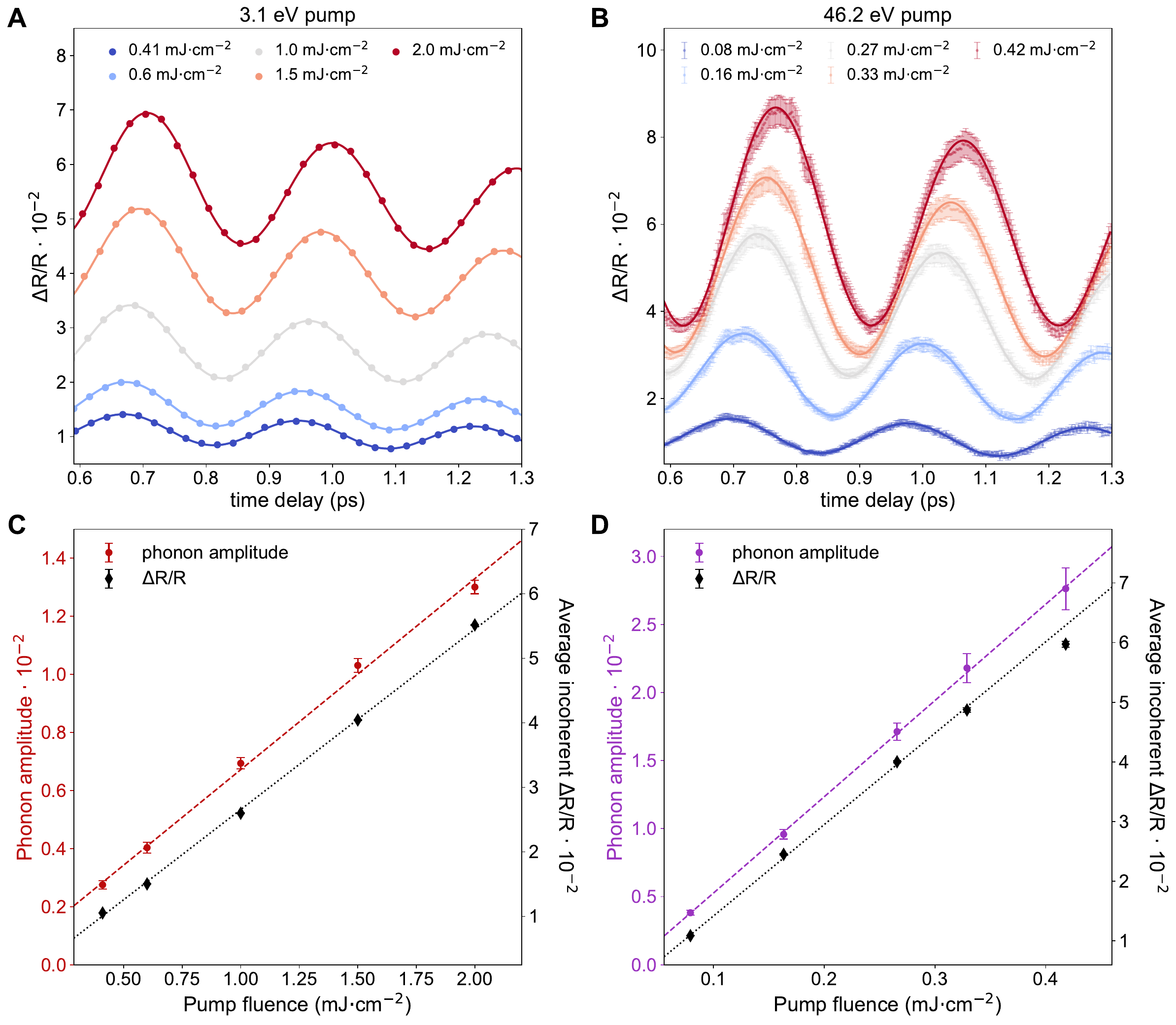}
    \caption{
    \textbf{Optical and XUV transient reflectivity fluence dependence of tellurium.} 
    (\textbf{A}) Transient reflectivity of tellurium as a function of the 3.10~eV pump fluence in the 0.59-1.3~ps delay range.
    (\textbf{B}) Transient reflectivity of tellurium as a function of the 46.2~eV pump fluence in the 0.59-1.3~ps delay range. The error bars are $\pm 2\sigma$ ($\sim$95\% confidence interval) of the total uncertainty of the transient reflectivity for each pixel.
    (\textbf{C}) Phonon amplitude (red dots) and average incoherent transient reflectivity signal (black diamonds) as a function of the 3.10~eV pump fluence and their linear fits (dashed and dotted lines, respectively).
    (\textbf{D}) Phonon amplitude (purple dots) and average incoherent transient reflectivity signal (black diamonds) as a function of the 46.2~eV pump fluence and their linear fits (dashed and dotted lines, respectively). The transient reflectivity starts deviating from linearity at the highest pump fluence. In (\textbf{C}) and (\textbf{D}) the error bars are $\pm 2\sigma$ ($\sim$95\% confidence interval) of the fitted parameters computed with the bootstrap method.
    }
    \label{fig:opt_vs_XUV_fluence}
% \vspace{0pt}
\end{figure}

The optical transient reflectivity experiment is performed as a function of the pump fluence in order to provide a direct comparison with the XUV traces reported in Figure~\ref{fig:fluence} of the main manuscript. The data are collected using a circularly polarized 3.10~eV pump and a linearly polarized 1.55~eV probe orthogonal to the \textit{c} axis of the tellurium crystal. The average optical traces are shown in Figure~\ref{fig:opt_vs_XUV_fluence}A in the 0.41-2.00~mJ/cm${}^2$ pump fluence range and fitted using equation~\eqref{eq:transients} of the main manuscript.
Figure~\ref{fig:opt_vs_XUV_fluence}B reports the XUV traces for the pump fluences 0.08-0.42~mJ/cm${}^2$, which cover a range of transient reflectivity amplitudes similar to the optical transients. The fit parameters for the optical and XUV traces are shown in Figure~\ref{fig:opt_vs_XUV_fluence}C,D, respectively. Both phonon amplitudes and average transient reflectivity scale linearly with the pump fluence, except at the highest XUV pump fluence, where deviations from the linear trend start to be observed. This linearity makes their mutual relationship insensitive to the absolute fluence calibration or photocarrier density (see section~\ref{SM:XUV_photocarrier_density}), and the coherent phonon driving efficiency can be expressed as the phonon amplitude against the average transient reflectivity signal (incoherent component), providing an internal reference for the measurements. 
The slopes of the resulting curves, which are reported in Figure~\ref{fig:fluence}C of the main manuscript, are equal to \mbox{0.46 $\pm$ 0.02} and \mbox{0.23 $\pm$ 0.01} for the 46.2~eV and 3.10~eV pump photon energies, respectively, confirming the stronger phonon activation of tellurium under X-ray light compared to optical pulses.

\section{XUV photocarrier density}\label{SM:XUV_photocarrier_density}
The XUV photogenerated carrier density is computed with the Beer-Lambert equation:

\begin{equation}
    n(z)=\eta\,\frac{(1-R)\,F}{\hbar\omega\,\delta_{\rm pump}}\,e^{-z/\delta_{\rm pump}}
\end{equation}

with $F$ the incident peak fluence, $\delta_{pump}$ the $1/e$ intensity penetration depth, $R$ the
pump reflectivity (close to zero in the XUV range) and $\eta$ the number of electron–hole pairs created per absorbed photon.
The large mismatch between the 46.2~eV pump and 1.58~eV probe intensity penetration depths in tellurium, respectively 310~nm \cite{henke1993x} and 42~nm \cite{tutihasi_optical_1969}, makes the probe volume uniformly pumped at 88\% of the incident value of the XUV density, as the pump profile marginally decays across the probed slab.
The $\eta$ factor accounts for photomultiplication processes that follow the energy dissipation of the initial XUV photon energy deposited in the system, such as Auger decay and impact ionisation \cite{medvedev2010transient,bohinc2019nonlinear}.

In order to estimate the $\eta$ value, we follow the relations for the average electron-hole pair creation energy given in Ref. \cite{klein1968bandgap}:

\begin{equation}
    E_{\rm pair}\simeq 2.8\,E_{\rm gap}+(0.5\text{–}1.0)\ \mathrm{eV},\qquad \eta=\hbar\omega/E_{\rm pair}
\end{equation}

and obtain $\eta$ between 24 and 32 for Te, whose band gap is ~0.33~eV \cite{tutihasi_optical_1969}. The resulting photocarrier density ranges from $7.3\cdot10^{18}-9.9\cdot10^{18}$ $\mathrm{cm^{-3}}$ up to $3.9\cdot10^{19}-5.2\cdot10^{19}$ $\mathrm{cm^{-3}}$ for XUV fluences of 0.08-0.42~mJ/cm${}^2$. 
These values are close to the excitation range of the 3.10~eV pump, which has a penetration depth of 12.5~nm (absorbed up to 23\% of the incident intensity across the probe volume), a reflectivity at normal incidence of 0.48 and $\eta$ values of 1.6-2.2, leading to $1.3\cdot10^{20}-1.7\cdot10^{20}$ $\mathrm{cm^{-3}}$ and $6.2\cdot10^{20}-8.4\cdot10^{20}$ $\mathrm{cm^{-3}}$ for fluences of 0.41-2.00 ~mJ/cm${}^2$. We remark that, while the nominal photocarrier densities of the two experiments differ by one order of magnitude, these numbers inherit the full uncertainty of two independent fluence calibrations, of the probe-weighting treatment and of the estimated photomultiplication factor. An absolute comparison between large-scale facility and table-top measurements therefore remains a complex challenge. This concern provides the strongest support to the internal calibration adopted in the main manuscript for the definition of the phonon driving efficiency.

%%%%%%%%%%%%%%%% SUPPLEMENTARY FIGURES %%%%%%%%%%%%%%%

% Figure: XAS

\clearpage 

% Figure: crocodile

\clearpage 

% Figure: optical 2D map

\clearpage 

% Figure: fluence dependence comparison

\clearpage 

\end{document}